# Test of detuning system for dielectronic recombination experiment at CSRm

MENG Ling-Jie(孟令杰)[1,2 ;1)] MA Xin-Wen(马新文)[1;2)] V. V. Parkhomchuk[3] YANG Xiao-Dong(杨晓东)[1] V. B. Reva[3] LI Jie(李杰)[1] MAO Li-Jun(冒立军)[1] MA Xiao-Ming(马晓明)[1] YAN Tai-Lai(晏太来)[1] XIA Jia-Wen(夏佳文)[1] YUAN You-Jin(原有进)[1] XU Hu-Shan(徐瑚珊)[1] YANG Jian-Cheng(杨建成)[1] XIAO Guo-Qing(肖国青)[1]

[1]Institute of Modern Physics, Chinese Academy of Sciences, Lanzhou, 730000, China

[2]Graduate University of Chinese Academy of Sciences, Beijing, 100049,China

[3]Budker Institute of Nuclear Physics, Laverentyeva 11, 630090 Novosibirsk, Russia

**Abstract:** The storage ring equipped with an electron cooler is an ideal platform for dielectronic recombination (DR) experiments. In order to fulfil the requirement of DR measurements at the main Cooler Storage Ring, a detuning system for the precision control of the relative energy between the ion beam and the electron beam has been installed on the electron cooler device. The test run using 7.0 MeV/u $C^{6+}$ beam was performed to examine the influence of this system on the performance of the stored ion beam. The Schottky spectra and the ion beam currents were recorded to monitor the beam status. The influence of pulse heights and widths of the detuning voltage on the ion beam was investigated. For the small pulse height, the experimental results from the Schottky spectrum were in good agreement with the theoretical results. The frequency shift in the Schottky spectrum is significantly reduced for the short pulse width. For the large pulse height, an oscillation phenomenon was observed. From the Schottky spectrum, we found the oscillation amplitude is dependent on the pulse width of detuning and the ion beam intensity. The detailed description of the phenomenon and the theoretical model based on the plasma oscillation was discussed in this paper.

## 1 Introduction

Dielectronic recombination (DR) is a fundamental electron-ion collision process [1], and is of great importance in various plasmas [2, 3]. It determines the charge-state distribution of atoms in ionized gases and also the spectrum of electromagnetic radiation emitted by such a gas. DR cross sections and rate coefficients are required for diagnosing the status of plasma. In the past few decades, many experiments have been conducted in an effort to acquire DR rates. Early electron-ion recombination experiments were carried out by applying plasma techniques such as theta-pinch [4]. Since the ion charge states accessible to these experiments were rather limited at the beginning, other experimental access to DR cross sections was pursued. The crossed- and merged-beam measurements of free electrons and ions had made vast progresses [5-7]. However, all of these methods have similar defects such as low counting rate, very limited energy resolution and high background.

The use of an electron cooler at storage rings [8, 9] for recombination studies has many advantages compared to traditional methods [10], such as high luminosity due to the high revolution frequency of ion beams, very low background because of ultra-high vacuum, and Phase-space cooling which improves the energy resolution in the experiments, etc. Such kind of experiments has been carried out at several ion storage ring facilities, e.g. TSR in Heidelberg [11-16], ESR in Darmstadt [17], and CRYRING in Stockholm [18, 19]. On one hand, fruitful results on the structure of few electron ions and QED effects have been obtained by the DR spectroscopy [12, 13, 14, 17] on these facilities. On the other hand, absolute recombination rate coefficients for astrophysical and other plasma physical applications were determined [15, 16, 18, 19].

In recent years, efforts have been made to achieve very high energy resolution in the recombination experiments at TSR. There, besides the electron cooler, a new electron beam device with photo-electron cathode has been installed and used as electron target in recombination experiments [20]. The ultra-cold electron target allows dielectronic recombination experiments with extremely high energy resolution, even the hyperfine structure has been observed [14]. In combination with fragment separator, DR technique has been applied to measure the isotope shift of radioactive ions at ESR in GSI [17]. So, DR is now developed as an excellent tool, not only to determine atomic parameters in electron-ion interaction, but also to study the nuclear size and spin for radioactive species.

In the end of 2007, the construction of the cooler storage ring (CSR) was completed and the commissioning was successful [21]. Later, a test experiment for detection of the recombined ions was performed at the CSRm [22], proved that the measurement of electron-ion recombination at the CSRm is feasible. In order to well control the relative energy in DR measurements at the CSRm, a detuning system developed in cooperation with the Budker Institute of Nuclear Physics (BINP) was installed on the electron cooler device [23] of the CSRm. The details of the detuning system and the test experimental results will be reported in this paper.

1. **Detuning system and experiment**

The working principle of the fast modulation of the electron beam energy is based on idea of two power supply connected in series. The 35 kV power supply (PS) produces the constant voltage for the accelerating of the electron beam to the fixed energy. The value of pulse height relative to the cooling point can be changed with help of the independent power supply. The ±3 kV switch PS produces the fast switching between two values of the electron energy and realizes the detuning energy of the electron beam. The detuning system consists of two subsystems (see Fig.1). One locates in high-voltage terminal (HVT) and modulates the energy of the electron beam. The other is located at the ground potential (ST) and modulates the potential of the electrostatic plates. The electrostatic plates realize the electrostatic bending of the electron beam, therefore the impulses on HVT and plates should be synchronized with high accuracy to keep the position unchanged of the electron beam at the interaction zone. The waveform generator produces the synchronizing signal for both systems, and before and after each positive or negative pulse, the waveform generator will output separate standard NIM signals for data analysis, see the description of Fig. 1.The waveform of the pulsed voltage is a rectangular shape with adjustable amplitude, width and polarity. The rising time for the pulse is about 50 microseconds at 400V pulse voltage (see Fig.2). The minimum step size of the pulse height is 1V. The pulse width and the time between two pulses can be adjusted from 1 millisecond to 2500 milliseconds.

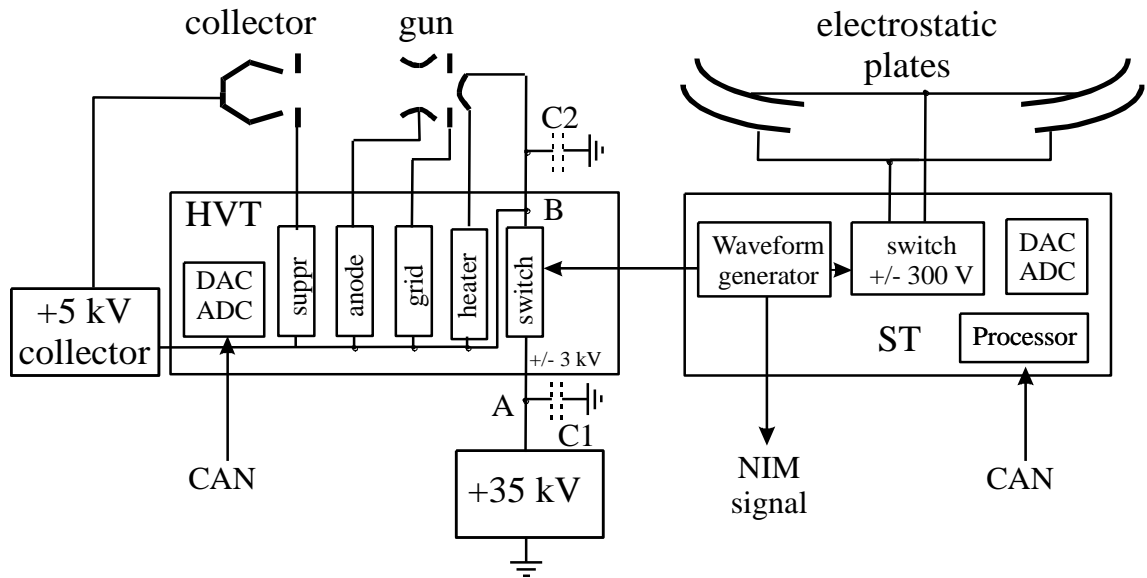

**Fig.1** Detuning system for fast changing of the electron beam energy. The high voltage terminal contents all power supplies for the electron beam operation. The heater PS produces cathode heating. The grid and anode PS control the shape of the electron beam and the electron beam current. The suppressor PS (suppr) controls the collector efficiency and the collector PS absorbs energy of the electron beam. The switch power supply forms the fast change of the electron beam energy. The ST module controls the voltage of the electrostatic plates for its bending in the toroid part of the magnetic field.

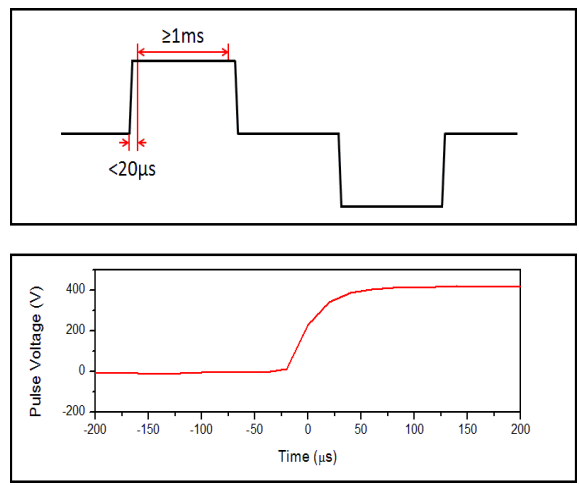

**Fig.2** The upper panel is a designed waveform of the detuning voltage. The lower panel is the measured signal from the output of the detuning system. The rising time is about 50μs at 400V pulse voltage.

The test experiment was performed at the electron cooler section of the CSRm. The layout of the CSRm is shown in Fig.3. A beam of 7.0MeV/u $C^{6+}$ was provided by the Sector Focusing Cyclotron (SFC) and injected into the cooler storage ring. The electron beam in the cooler was guided by a longitudinal magnetic field and collinearly overlapped with the ions in the cooling section with a length of 4 meters. The electron density was set about $1.3*10^7 cm^{-3}$ in the present measurements.

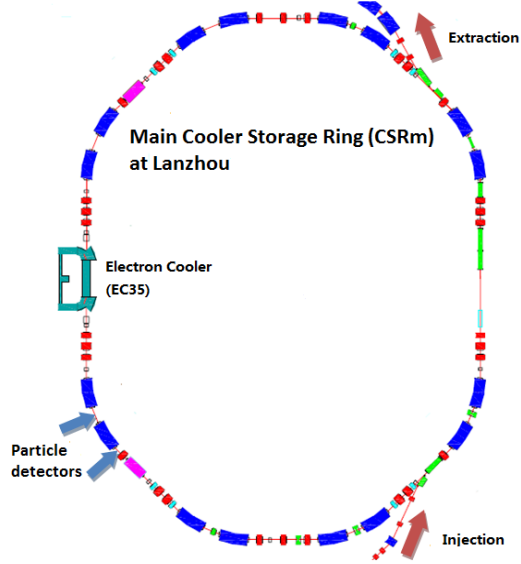

**Fig.3** Layout of the CSRm storage ring. In DR measurements, the circulating ion beam is merged with electron beam in the cooling section (EC35). After the downstream dipole magnets, recombined ions are separated from the primary ion beam and detected by the particle detectors.

To test the performance of the system, the electron energy was detuned away from cooling point by positive pulses during the measurements. The ion beam will be dragged by the electrons when the ion beam velocity is different from the electron beam velocity. The drag effect is caused by the cooling force and varies depending on the parameter settings, e. g. pulse height and pulse width. In storage ring, the change of ion's revolution frequency which is in direct proportion to the ion beam velocity can be monitored by the Schottky pick-up device. The signals from the Schottky device were recorded and analysed by a Tektronix RSA3408A spectrum analyser. The principles of Schottky spectrometry have been described elsewhere [24-26]. A DC Current Transformer (DCCT) was used to monitor the pimary ion beam current. A data collecting card (National Instruments USB-6210) combined with the commercial software (LabView) were used for data acquisition of the signals from the DCCT.

2.  **Results and Discussion**

The Schottky spectra which were recorded during detuning are shown in Fig. 4, where horizontal and vertical axes of each spectrum are frequency and time, respectively. The pulse width for each voltage was set to 100 ms. In the upper panel of the figure, the data were recorded for the pulse heights of (a) 0V, (b) 5V and (c) 10V. At the cooling point (0V), the frequency of the ion beam always remained the same, which indicates the detuning system itself has no influence on the ion beam. At the pulse height of 5V, the frequency of the ion beam was shifted (see peaks in the figure) due to the pulses. This indicates that the energy of the ion beam was changed by the detuned electron beam. At the pulse height of 10V, the largest shift in the measurements was observed. In the lower part of the figure, the data were recorded for the pulse heights of (d) 20V, (e) 30V and (f) 40V. In this range, the shift of the frequency decreases with the increase of the pulse height. The observed phenomena can be explained by the electron cooling force. The cooling force $F$ increases linearly with the increase of relative velocity between ions and electrons until it reached its maximum value, and then decreases as $F \propto 1/(V_e - V_i)^2$, where $V_e$ and $V_i$ is velocity of electrons and ions respectively.

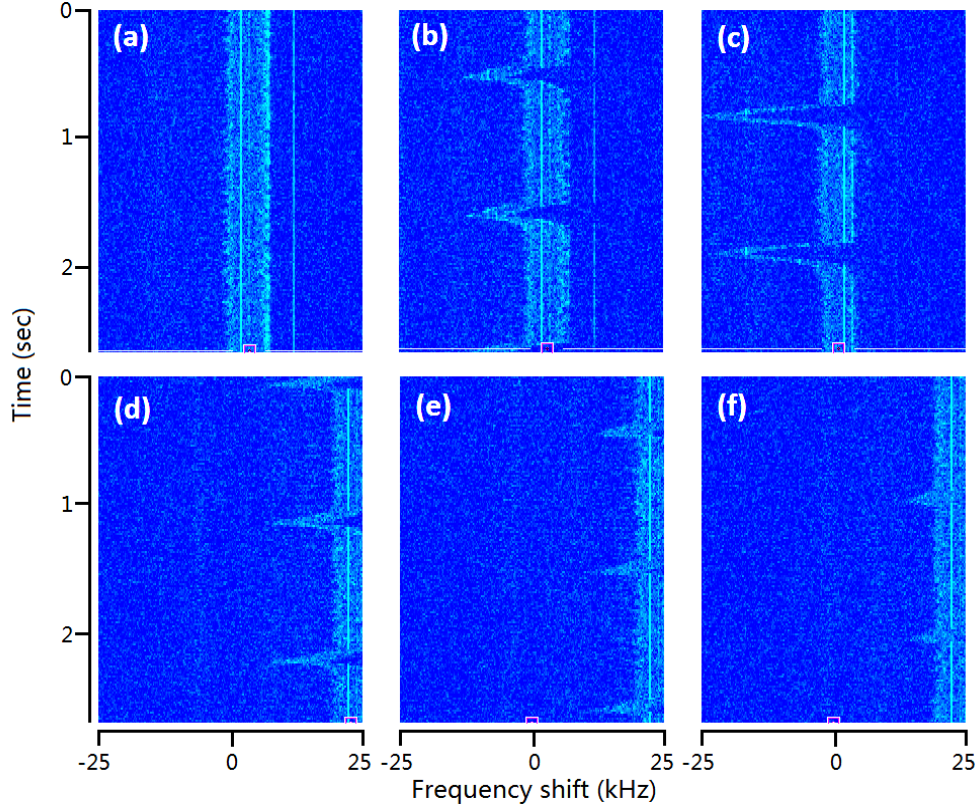

**Fig.4** Schottky spectrum recorded with pulse heights varying from 0 to 40V. The pulse width was 100ms and interval between two pulses was 1s. The central frequency for the upper part and the lower part of the figure are 17.988 MHz and 17.968 MHz, respectively. The span of the spectrum is 50 kHz.

According to the parameters used in the measurements, the frequency shift has been calculated. When the velocity of the electrons is detuned, the ions will have acceleration

$$a(V_i, V_e) = -a_0 \cdot \frac{V_i - V_e}{\left[(V_i - V_e)^2 + V_{eff}^2\right]^{\frac{3}{2}}} \tag{1}$$

with

$$a_0 = 4 r_e r_i n_e c^4 \eta_c \ln\left(\frac{\rho_{max}}{\rho_L}\right) \tag{2}$$

where $V_{eff}$ is the effective spread of the electrons velocity, $n_e$ is the number density of the electron, $c$ is the speed of light, $\eta_c = 0.025$ is the ratio between the length of interaction zone and the ring circumference. $\rho_{max}$ is the maximal impact parameter, $\rho_L$ is the radius of Larmor rotation at the magnet field in the cooling section, $r_e$ and $r_i$ are classical radius of electron and ion, respectively. The initial velocity variation of the ions is zero, we have

$$\Delta V_{i0} = 0 \tag{3}$$

The velocity variation of the ions at any intermediate time can be written as

$$\Delta V_{i(k+1)} = \Delta V_{ik} + a(V_{ik}, V_{ek}) \cdot \Delta t \quad (4)$$

The frequency shift in the Schottky spectrum is

$$\Delta f_k = f_0 \eta_p \frac{\Delta V_{ik}}{c\beta} \quad (5)$$

with

$$\eta_p = \frac{1}{\gamma^2} - \frac{1}{\gamma_t^2} \quad (6)$$

Where $f_0$ is the central frequency of the Schottky spectrum, $\eta_p$ is the mean frequency dispersion function, $\gamma$ is the relativistic Lorentz factor, $\gamma_t$ is the transition point of the storage ring.

The calculated results are shown in Fig.5 corresponding to the parameters of measurements in Fig.4. By comparing the figures 4 and 5, it can be found that the relative shift of the frequencies is in good agreement with the experimental results, and the absolute value of the frequency shifts is about the 90% of the experimental values. This indicates that the detuning system is working properly for this region of detuning voltages.

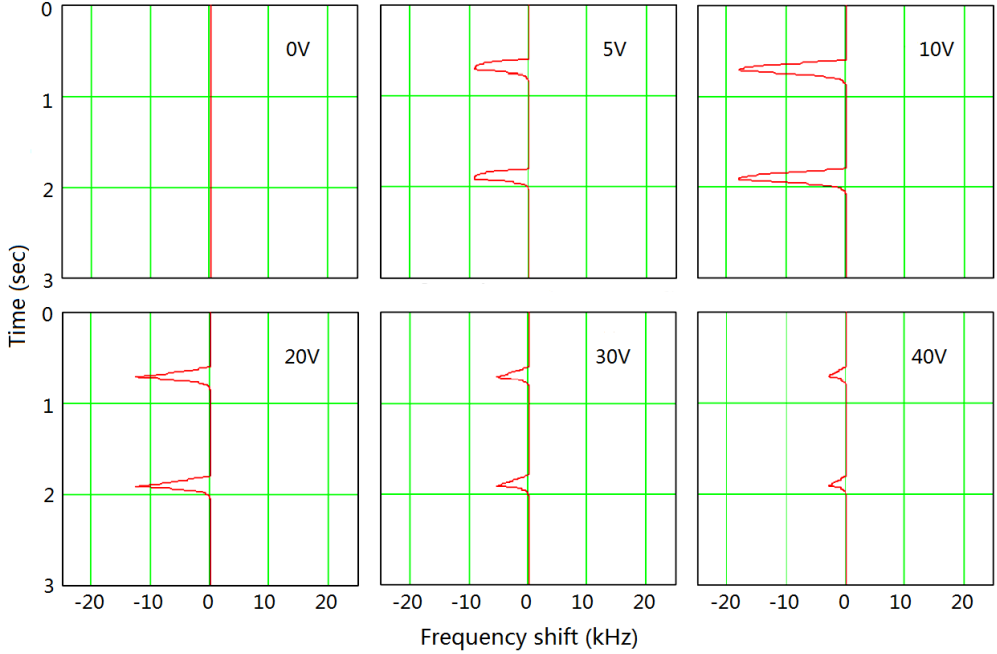

**Fig.5** Theoretical results of frequency shift under various detuning conditions calculated according to the parameters of measurements in Fig.4. Horizontal and vertical axes are frequency and time, respectively.

To study the influence of the pulse width on the ion beam during detuning, the Schottky spectrum for pulses width of 100ms and 20ms are compared. The pulse height is 10V which produces the most predominant shift in the measurements. The test results are shown in Fig.6. The frequency shift is significantly reduced for the pulse width of 20ms (right panel). Therefore, the influence of detuning on the ion beam can be neglected when the pulse width is small enough.

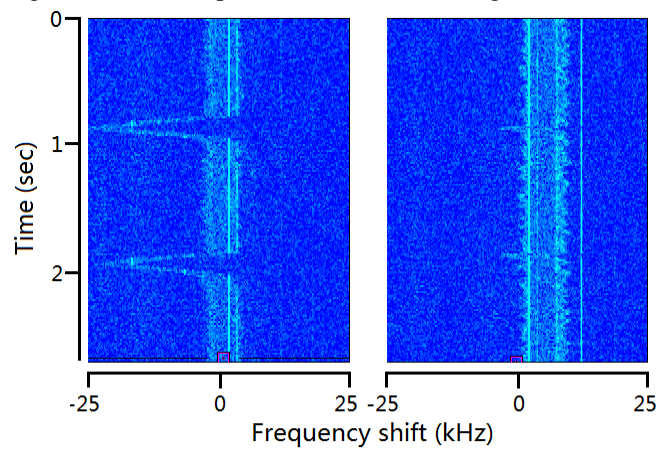

**Fig.6** Schottky spectra recorded for pulse width of 100ms (left) and 20ms (right), respectively. The pulse height was 10V and the interval was 1s. The central frequency for the left panel and the right panel is 17.988 MHz. The span of the spectrum is 50 kHz.

Generally, the required small relative energy in the centre of mass frame corresponds to a large pulse height of detuning in the DR experiments, e.g. 50 eV relative energy corresponding to the pulse height of about 900V for 7.0 MeV/u ion beam. The test of large pulse height was carried out up to 1000V. In principle, the cooling force can be neglected when the relative energy is large. Therefore, the velocity of the ion beam should not be significantly changed for large pulses. Unexpectedly, oscillations in the Schottky spectrum were observed when the large pulse heights were applied. A typical result is shown in Fig.7 where the pulse height is 700V. The pulse width from left to right is 100ms and 20ms, respectively. The observed oscillations are weaker for the pulse width of 20ms. This suggests that using short pulse width during detuning is an effective way to reduce the oscillation level when the DR experiment is performed.

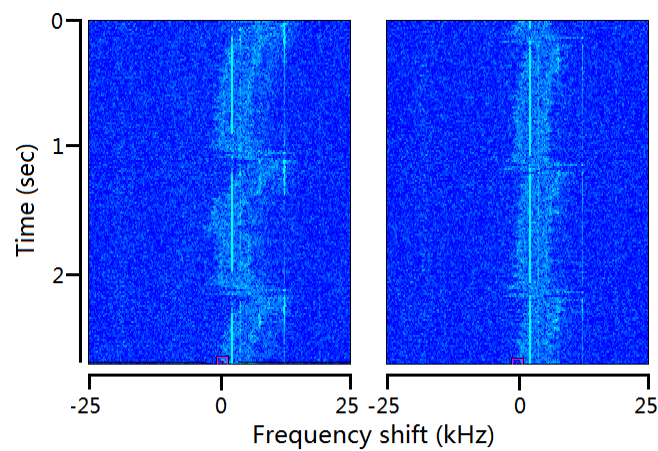

**Fig.7** Recorded Schottky spectrum. The pulse height is 700V and the interval is 1s for the pulse width of 100ms (left) and 20ms (right), respectively. The central frequency for the upper part and the lower part of the figure is 17.988 MHz. The span of the spectrum is 50 kHz.

To further study the phenomenon, the influence of large pulse height on the ion beam intensity was investigated. The ion beam intensities versus stored time for pulse heights of 100V, 200V and 400V are shown in Fig.8. The vertical axis is logarithmic. In order to facilitate the comparison, the initial intensities are normalized to the same value. For the pulse heights of 100V and 200V, the intensities decrease linearly and the loss rates (which are related to the slope of the line) are small. For the pulse height of 400V, very high loss rate of the ion beam intensity were observed. Furthermore, we found that the loss rate of the ion beam decreased when the ion beam current was small, comparison to a straight line (dash line in the figure). The result shows that the oscillation which caused the rapid beam loss is dependent on the ion beam intensity.

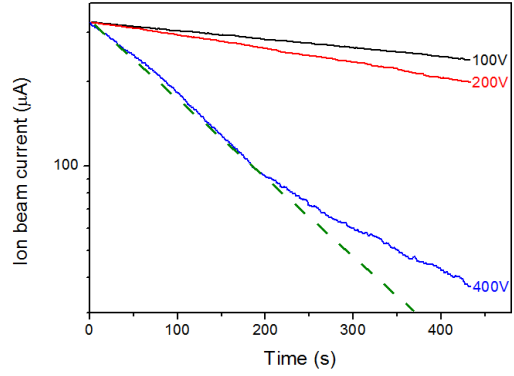

**Fig.8** The ion beam intensity vs. time for the pulse height of 100V, 200V and 400V. The pulse width is 20ms and the interval between two pulses is 100ms. The vertical axis is logarithmic. The dashed line is only for guidance.

Similar phenomenon was also observed at CELSIUS and was called "electron heating" by Reistad [27]. The lifetime of a bunched 48MeV proton beam typically changes from 50-100s without electron cooling to 0.5-1s when it is exposed to 100mA electrons with energy at the cooling point. Furthermore, when the electron beam energy was detuned far away from the cooling point, the lifetime of the proton beam became even shorter. The author considered the phenomenon as a single-particle instability which may be due to excitation of resonances by non-linear electrical fields from the electron beam.

To explain the electron heating phenomenon, Parkhomchuk proposed another hypothesis [28]. The theoretical model to describe the electron heating is based on two beam plasma oscillation. In the cooling section, the equation of motion for plasma oscillations can be written as:

$$\frac{d^2 x_e}{dt^2} = -\frac{e}{m} E_p \qquad (7)$$

$$\frac{d^2 x_i}{dt^2} = -\frac{Ze}{M} E_p \qquad (8)$$

With

$$E_p = 4\pi e (n_e x_e - n_i x_i) \qquad (9)$$

Where $E_p$ is the electric field of plasma oscillations acting on both beams, $e$ is the electric charge, $Z$ is the atomic number, $m$ and $M$ are the effective mass of the electron and the ion, $n_i$ is the number density of the electron and ion beams, respectively. The oscillation equation can be written in the form

$$\frac{d^2 E_p}{dt^2} = -\left(\omega_e^2 + \omega_i^2\right) E_p = -\omega_p^2 E_p \tag{10}$$

Where $\omega_e$ and $\omega_i$ are plasma frequency of electron and ion beams respectively. The ion beam runs many turns at the storage ring and interacts with fresh electron beam at each turn. Therefore, at the entrance of the cooling section the ion beam has some coordinates $x_i(0)$ and initial velocity $dx_i(0)/dt$, the electron beam has $x_e(0) = 0$ and $dx_e(0)/dt = 0$.

The self-consistent solution can be written as:

$$\begin{pmatrix} x_i \\ dx_i/dt \end{pmatrix} = A(t) \times \begin{pmatrix} x_i \\ dx_i/dt \end{pmatrix}_0 \tag{11}$$

With specific initial conditions, the determinant (det) of the matrix $A(t)$ can be calculated

$$|A| = 1 - 2\frac{\omega_e^2 \omega_i^2}{\omega_p^4}\left(1 - \cos(\omega_p \tau)\right) + \frac{\omega_e^2 \omega_i^2 \tau}{\omega_p^3}\sin(\omega_p \tau) \tag{12}$$

Where $\tau$ is the time of interaction in the cooling section (in the ion beam system).

In Fig.9 we show the calculated results of det(A) for $C^{6+}$ ion beam with number density of $10^5 cm^{-3}$ (solid line) and $10^6 cm^{-3}$ (dash line), respectively. The value of det(A) determined the energy transfer between two beams. If det(A) < 1, the electrons absorb energy from the ions, indicating fast cooling of the ion beam. If det(A) > 1, the electrons release energy, indicating fast heating of the ion beam. Furthermore, the heating effect is not obvious when the ion density is small; this is in agreement with what we discussed earlier.

It should be noticed that the heating effect only appears when the electron energy is detuned in the CSRm, and the theory was developed for description of the instability at zero velocity difference. Therefore, the factor provoked the instability at high velocity difference is still not clear. The detailed model of this plasma oscillation and its production mechanism are needed to gain a closer understanding of the phenomenon.

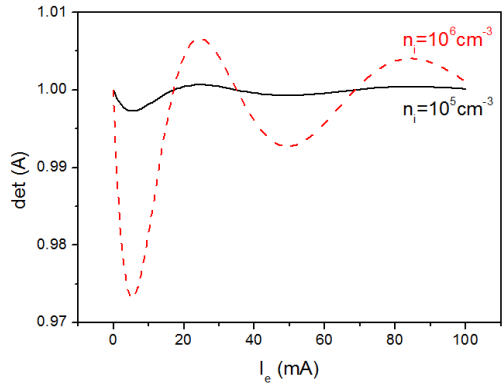

**Fig.9** Variation of matrix determinant vs. electron beam current for $C^{6+}$ ion beam with number density of $10^5 cm^{-3}$ (solid line) and $10^6 cm^{-3}$ (dash line), respectively.

3. **Conclusions**

The detuning system which varies the relative energy between the ion beam and the electron beam was installed at the CSRm for performing the DR experiment. The system was tested by varying pulse height and pulse width using stored, 7.0 MeV/u $C^{6+}$ ion beam. For small pulse heights, the frequency shifts in the measured Schottky spectrums are very close to the results of theoretical values. To test the influence of pulse width on the ion beam quality, the results for the pulse width of 100ms and 20ms with the pulse height of 10V are compared. The frequency shift was significantly reduced for the pulse width of 20ms. For large pulse heights, oscillations were observed in the Schottky spectrums. According to the test results, the oscillation amplitude is dependent on the pulse width and the ion beam current. The theoretical model which is based on the plasma oscillation was employed to explain the phenomenon. The results of the theoretical model show that the heating effect has a strong dependence on the density of the electron beam and the ion beam. More test experiments are required to gain a better understanding of the heating effects at large pulse heights.

With the newly installed detuning system, the test run proves that the DR measurements are feasible at the CSRm. The results discussed in this paper will be helpful for the upcoming DR experiment which is in preparation.

*MENG Ling-Jie would like to thank Wen Wei-Qiang for his help in English improvement.*